\documentclass[%
 reprint,
superscriptaddress,
 amsmath,amssymb,
 aps,
]{revtex4-2}

\usepackage{graphicx}
\usepackage{dcolumn}
\usepackage{bm}

\usepackage[version=3]{mhchem} 
\usepackage{amsfonts}
\usepackage{xcolor}
\usepackage{soul}

\begin{document}

\title{The inverse Faraday effect at Mie resonances}
\preprint{APS/123-QED}

\author{Denis M. Krichevsky}
\email{krichevskii.dm@phystech.edu}
\affiliation{Russian Quantum Center, Moscow 121205, Russia}
\affiliation{Moscow Institute of Physics and Technology (National Research University), Dolgoprudny 141700, Russia}

\author{Daria O. Ignatyeva}
\email{daria.ignatyeva@gmail.com}
\affiliation{Russian Quantum Center, Moscow 121205, Russia}
\affiliation{Faculty of Physics, M.V. Lomonosov Moscow State University, Moscow 119991, Russia}

\author{Vladimir I. Belotelov}
\affiliation{Russian Quantum Center, Moscow 121205, Russia}
\affiliation{Faculty of Physics, M.V. Lomonosov Moscow State University, Moscow 119991, Russia}



\date{\today}

\begin{abstract}

Nowadays, dielectric nanophotonics enables almost lossless resonant interaction between light and matter at the nanoscale. We show both theoretically and by electromagnetic simulations, that the peculiar nature of Mie-resonance induced effective magnetic fields contrasts sharply with the optomagnetism of smooth bulk materials. Mie resonances produce strongly nonuniform effective magnetic fields generated by the inverse Faraday effect. Different orders of optical resonances are characterized by different types of the effective magnetic field patterns. The number of point-like inverse Faraday effect sources can be controlled by adjusting the pump wavelength, allowing for the selective launch of spin waves with submicron wavelengths. A distinguish feature of the Mie-resonance-induced effective magnetic fields is the vortex structure that can be used for the magnetic skyrmion generation. The proposed approach considerably broadens the scope of nanoscale optomagnetism.

\end{abstract}

\maketitle

\section*{Introduction}
Interaction of ultrashort laser pulses with magnetically ordered materials yields breakthrough technologies such as nonthermal magnetic recording \cite{stupakiewicz2017ultrafast,stupakiewicz2019selection}, magnetic order manipulation in antiferromagnets \cite{manz2016reversible}, subpicosecond spin dynamics excitation \cite{bossini2016macrospin} and coherent spin wave manipulation \cite{hortensius2021coherent, satoh2012directional}. To control laser-induced processes in magnetic materials at the nanoscale nanostructures are employed~\cite{ignatyeva2022all, qin2022nanophotonic}. One of the most wide-spread nanostructures are plasmonic. They possess high electromagnetic field localization and enhancement caused by excitation conducting electrons primarily in noble metals \cite{maier2007plasmonics}. Such excitation at the metal-dielectric interface can form propagating or localized plasmonic surface modes of electromagnetic nature resulting in electromagnetic field enhancement within several hundreds of nanometers. Surface plasmons were found to be efficient for spin dynamics control at optical and terahertz frequencies~\cite{schlauderer2019temporal,kolodny2019resonant,ignatyeva2019plasmonic}. However, metals that support the excitation of plasmons lack high light absorption at optical frequencies leading to heating of the metals themselves and surrounding media \cite{jauffred2019plasmonic}.

To overcome this problem dielectric structures are about to replace their metallic counterparts in nanophotonic devices \cite{baranov2017all}. Until now all-dielectric nanostructures have been found promising not only for linear and nonlinear optical applications \cite{zograf2021all,decker2016resonant,koshelev2020subwavelength,krasnok2015towards}, but also for advanced magneto-optics and optomagnetism \cite{voronov2020magneto,voronov2020resonances,ignatyeva2020all,ignatyeva2022magneto,kimel20222022,chernov2020all,ignatyeva2022all,wojewoda2023observing}. The latter one uses inverse magnetooptical effects, such as inverse Faraday or inverse Coutton-Mouton effects, to stimulate spin dynamics in magnetic materials. The effects result in effective magnetic field generation in transparent magnetic media \cite{pitaevskii1960electric, pershan1966theoretical}. Optical modes in dielectric nanostructures change effective magnetic field distribution and allows for standing spin waves launch\cite{krichevsky2021selective, chernov2020all,krichevsky2024spatially,ignatyeva2024optical}.

Among variety of optical modes in all-dielectric nanophotonics Mie resonances play a special role resulting in enhanced harmonics generation \cite{melik2017selective,smirnova2018multipolar}, nonradiating anapole states \cite{wei2016excitation,miroshnichenko2015nonradiating} and metalensing \cite{kruk2017functional}. The excitation of Mie resonances in all-dielectric nanoelements is accompanied with a specific distribution of the electromagnetic field inside the nanoelement~\cite{kuznetsov2016optically}. This feature is crucial for optomagnetic interaction in transparent magnetic media because it provides additional degree of freedom for magnetic order manipulation and spin waves control.

In this work we theoretically and by electromagnetic modeling demonstrate how optical Mie modes in all-dielectric bismuth iron garnet  nanospheres can be implemented for spatially inhomogeneous inverse Faraday effect (IFE) magnetic field generation. The degree of IFE field nonuniformity is found to be dependent on Mie modes type (magnetic and electric) and order (dipole, quadrupole and octupole). Furthermore, a peculiar character of the Mie modes is also responsible for IFE field vorticity inside the nanosphere and can be used for the skyrmion generation.

\section{Theoretical basis of IFE in nanoparticles}
\subsection{IFE in magnetic materials}
The inverse Faraday effect is an optomagnetic effect which reveals itself as optically-induced magnetic field generation in optically transparent media~\cite{kirilyuk2010ultrafast}. For the case of an isotropic magnetic medium or a cubic crystal~\cite{vivs1986magneto} that is the case of iron-garnets this effective magnetic field $\mathbf{H}^\mathrm{IFE}$ produced by an optical pulse is:

\begin{equation}
\label{eq:equation3}
\textbf{H}^\mathrm{IFE}=-\frac{g}{16 \pi M_s}\mathrm{Im}(\textbf{E} \times \textbf{E*})    
\end{equation}
where \(g\) is a gyration constant and \(M_s\) is a saturation magnetization. 

Thus, for the smooth films or crystals only circularly polarized incident light can induce IFE field. This field is directed along the incident light wavevector \cite{kirilyuk2010ultrafast}. 

The situation changes if the illuminated nanostructure supports optical modes and polarization of the incident light is transformed inside the nanostructure. For instance, guided TM modes of a planar waveguide or a plasmonic structure possess elliptical polarization (\(E_x\) and \(E_z\) components have \(\pi/2\) phase shift). Such modes are known to induce IFE field that is orthogonal to the mode propagation direction and the incident light wavevector~\cite{krichevsky2021selective, belotelov2012inverse}. Such manifestation of IFE is also called the inverse transverse magneto-optical Kerr effect due to the specific direction of the light-induced magnetic field. Thus, to determine the effective magnetic field induced by IFE it is necessary to examine the electric field distributions within a nanoparticle.

In this study, we focus on the distributions of the IFE and do not limit ourselves by fixing the direction of the external magnetic field and magnetization with respect to the light wavevector. This is done with the intention of analyzing the entire range of IFE patterns. In a specific configuration of the magnetization \(\mathbf{M}\) (e.g., along the light wavevector or perpendicular to it), a certain IFE component parallel to the magnetization will not produce excitation torque \(\mathbf{T}\propto\mathbf{M} \times \mathbf{H}^{IFE}\) whereas other IFE components will induce magnetization dynamics.

\subsection{Mie modes of a magnetic nanoparticle}

Localized Mie resonances in nanoparticles that are confined in all directions are expected to exhibit quite complicated behavior because they drastically change light polarization. Mie resonances are described by solving the light scattering
problem on a single spherical particle of arbitrary radius and refractive index \cite{mie1908beitrage}. The total scattering and extinction cross section can be expressed as a series of spherical multipolar terms. At optical frequencies magnetic permeability \(\mu=1\). Thus, of the electromagnetic energy concentrated inside a dielectric nanoparticle core can be written down as follows\cite{bohren2008absorption}:

\begin{eqnarray}
\label{eq:equation4}
    \mathbf{E}_\mathrm{int} = \sum E_q (c_q \mathbf{M}_{o1q}^{(1)}-i d_q \mathbf{N}_{e1q}^{(1)}),  \\
\label{eq:equation5}
    \mathbf{H}_\mathrm{int} = -n\sum E_q (d_q \mathbf{M}_{e1q}^{(1)}+i c_q \mathbf{N}_{o1q}^{(1)}), 
\end{eqnarray}

where \(\mathbf{E}_\mathrm{int}\) and \(\mathbf{H}_\mathrm{int}\) are electric and magnetic fields inside the nanoparticle, \(c_q(\lambda)\) and \(d_q(\lambda)\) are wavelength-dependent scattering coefficients, \(\mathbf{M}_{o(e)1q}^{(1)}\) and \(\mathbf{N}_{o(e)1q}^{(1)}\) are magnetic and electric vector spherical harmonics correspondingly, $n$ is refractive index of the sphere and \(E_q=i^q E_0 \frac{(2q+1)}{q(q+1)}\), $q$ is a natural number \cite{bohren2008absorption}. The letters $o$ and $e$ represent odd and even spherical harmonics, respectively. The maxima of \(c_n(\lambda)\) and \(d_n(\lambda)\) coefficients determine the spectral positions of the Mie resonances. Vector spherical harmonics are usually written in the following form:

\begin{eqnarray}
\label{eq:equation6}
  \mathbf{M}_{e1q}=\nabla \times (\mathbf{r}\psi_{e1q}) =  \nabla \times (\mathbf{r} [\cos(\varphi)P_q(\cos\theta)z_q])\\
\label{eq:equation7}
    \mathbf{N}_{e1q}=\nabla \times \mathbf{M}_{e1q}= \nabla \times\nabla \times (\mathbf{r} [\cos(\varphi)P_q(\cos\theta)z_q]\\
\label{eq:equation8}
\mathbf{M}_{o1q}=\nabla \times (\mathbf{r}\psi_{o1q}) =  \nabla \times (\mathbf{r} [\sin(\varphi)P_q(\cos\theta)z_q])\\
\label{eq:equation9}
\mathbf{N}_{o1q}=\nabla \times \mathbf{M}_{o1q}= \nabla \times\nabla \times (\mathbf{r} [\sin(\varphi)P_q(\cos\theta)z_q]
\end{eqnarray}

In Eqs. \eqref{eq:equation6}-\eqref{eq:equation9} $\mathbf{r}$ is a position vector, $\varphi$ and $\theta$ are polar coordinates, $P_q$ is an associated Legendre function and $z_q$ is spherical Bessel function\cite{bohren2008absorption}. 
\par We consider a spherical bismuth-substituted iron-garnet (BIG) nanoparticle of 400 nm diameter with refractive index \(n_\mathrm{BIG}=2.6+0.01i\) surrounded by air. Such permittivity value is quite common for iron garnets with high Bi concentration, namely for $\mathrm{Bi_3Fe_5O_{12}}$ at the long wavelength visible and near infrared  spectral range (600 - 1500 nm) \cite{tepper2003pulsed,levy2019faraday}. Furthermore, imaginary part of dielectric permittivity index is much smaller than real one, so the dispersion of $\varepsilon$ can be neglected \cite{fakhrul2019magneto,fakhrul2021high}. Thus, on the one hand BIG nanoparticles possess quite high refractive index and low optical losses. On the other hand, they are known as the materials with the enhanced magneto-optical activity \cite{prokopov2016epitaxial}. The analysis is provided for a spherical nanoparticle since it has the analytical solution of the scattering problem. At the same time, the features of the IFE fields in the nanoparticles of the different shapes would be qualitatively similar.

\begin{figure*}[htbp]
\centering
\includegraphics[width=1\textwidth]{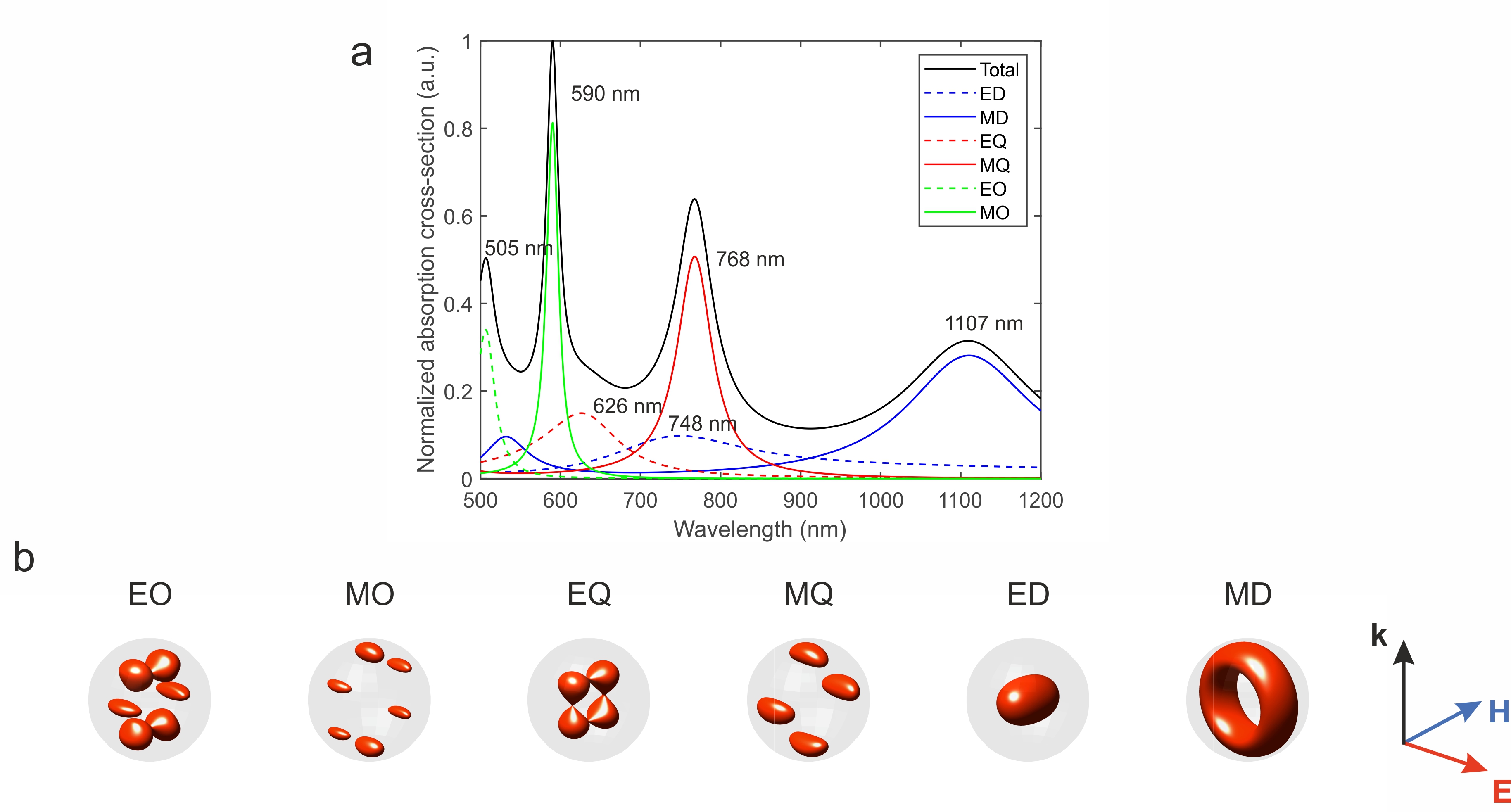} 

\caption{Optical properties of 400 nm BIG nanosphere. a - normalized absorption spectra obtained analytically on the bases of Mie theory. b - isosurfaces of the \(|E|^2\) inside the sphere at resonant wavelength. The letter "M" represent magnetic nature of the resonance, while "E" corresponds to electric one. Letters "D", "Q" and "O" designate dipole, quadrupole, octulope modes respectively.}
\label{fig: Modes_sphere}
\end{figure*}

Absorption spectra for the aforementioned nanoparticles under linearly polarized light illumination are presented in Figure~\ref{fig: Modes_sphere}a. The spectra consist of multiple peaks corresponding to Mie resonances excitation. For a single nanosphere the lowest resonances is magnetic dipole (MD) one which is typically observed at the lowest frequency (or highest wavelength). In our case MD resonances occur at \(\sim 1107\) nm. The lower the wavelength, the higher the order of excited Mie resonances. 

Figure~\ref{fig: Modes_sphere}a shows that some resonances (for example, MQ and ED) overlap. Both absorption energy (according to Joule's law) and IFE field are proportional to the light intensity within a nanoparticle. As a result, one can conclude that for the magnetic resonances $H^\mathrm{IFE}$ is at least twice as large as at electric resonances (Fig.~\ref{fig: Modes_sphere}a). In experiments Mie resonances spectrally overlap and cannot be selectively excited. Magnetic type resonances are the most intensive ones (see Fig.~\ref{fig: Modes_sphere}a). As a result, magnetic family of Mie modes primary contribute to the IFE effective fields generated in a nanosphere. At the same time, other configurations, such as other types of nanoparticles (e.g. core-shell structure) or other materials (such as magnetic semiconductors with high refractive index), may provide a different ratio of the electric and magnetic contributions to an absorption spectra and the electromagnetic fields. For this reason, further we will analyze the contribution of each of the resonances separately. Each of the Mie modes has unique distribution of the electromagnetic field profiles. Figure~\ref{fig: Modes_sphere}b shows isosurfaces of \(|E|^2\) distribution inside the particle core at corresponding resonant wavelength. The lowest order magnetic mode (MD) excited at $\sim1107$~nm posses \(|E|\) circulation~\cite{koshelev2020dielectric} of the electric field around $\mathbf{H}$ field. The higher the mode order, the more complicated \(|E|^2\) distribution becomes: the number of nodes and the light intensity hotspots grows. Moreover, with the increase of the mode number the polarization inside the particle becomes more and more twisted. This feature is critical for further investigation of the IFE fields.

\section{IFE hotspots as the confined spin dynamics sources}

\begin{figure*}[htbp]
\centering
\includegraphics[width=1\textwidth]{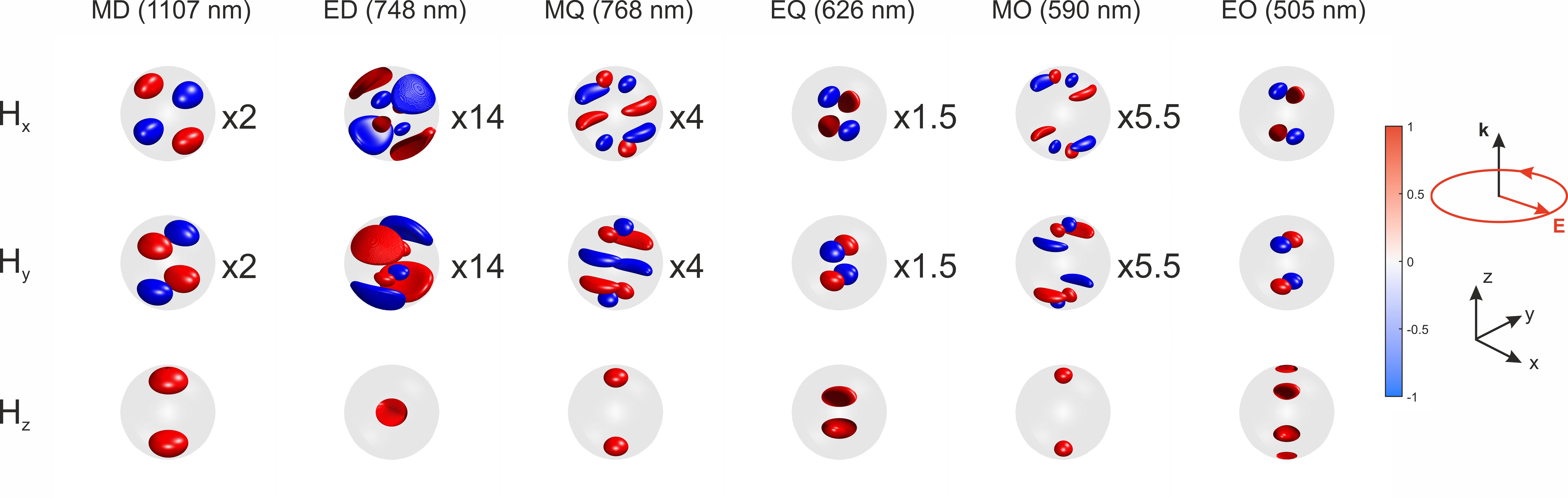} 
\caption{Isosurfaces of $H^\mathrm{IFE}$ field components distribution inside BIG nanosphere at the resonant wavelengths corresponding to the Mie modes excitation are shown at $\approx 1/2$ level of the maximum values. The numbers indicate the ratio between the $H^\mathrm{IFE}_{x}$, $H^\mathrm{IFE}_{y}$, and $H^\mathrm{IFE}_{z}$ components at the fixed wavelength.}
\label{fig: IFE_iso}
\end{figure*}

According to Eq.~\eqref{eq:equation3}, the cross product \(\textbf{H}^\mathrm{IFE} \propto \textbf{E} \times \textbf{E*}\) governs the IFE magnetic field within a nanosphere. It means that due to the inhomogeneity of the electromagnetic field inside a particle, all three non-zero components of the IFE field are expected. We investigate the IFE field generated in the nanoparticle by optical pulse with circular polarization at the wavelengths corresponding to the excitation of the Mie resonances. The light propagates along OZ direction (\(\mathbf{k} \parallel \)~OZ, see Figure \ref{fig: IFE_iso}) and its electric field components rotates in x-y plane. According to Eq.~\eqref{eq:equation3}, for a smooth BIG film such an excitation produces the only component of \(\textbf{H}^\mathrm{IFE}\  \parallel OZ\). For a BIG nanoparticle, the IFE pattern is completely different.

The $\mathbf{H}^\mathrm{IFE}$ field components for the observed resonances are presented in Fig.~\ref{fig: IFE_iso} as isosurfaces taken at $\approx 1/2$ level of the maximum value. Thus, areas marked by color represent the areas with the high magnitude of a certain ${H}^\mathrm{IFE}_j$ component, while the color (red or blue, respectively) denotes its sign, i.e. its direction. All resonances have a nonuniform distribution of the induced magnetic field components. The degree on nonuniformity increases with an increase of a Mie resonance order. 

Actually, only ED mode has \(H^\mathrm{IFE}_{z}\) that significantly dominates ($H^\mathrm{IFE}_{x}$, and $H^\mathrm{IFE}_{y}$). Other electric modes (EQ and EO, namely) have nearly the same magnitude of $H^\mathrm{IFE}_{x}$, $H^\mathrm{IFE}_{y}$, and $H^\mathrm{IFE}_{z}$ components. Thus, the induced magnetic field is inhomogeneous both in terms of localization and direction within a nanoparticle. For the magnetic family of modes, the magnitude of $H^\mathrm{IFE}_{x}$, and $H^\mathrm{IFE}_{y}$ components is several times smaller than $H^\mathrm{IFE}_{z}$. This difference increases with the increase of the mode number (from MD to MQ).

All of the Mie resonances are characterized by a quite peculiar distribution of $\mathbf{H}^\mathrm{IFE}$ within the nanoparticle (Fig.~\ref{fig: IFE_iso}). While the resulting direction of the generated IFE field will be discussed later, let us first analyze the volume and location of the IFE field hotspots inside a nanoparticle.

A mode volume, which describes the confinement of the cavity inner electromagnetic radiation, is an important resonator characteristics. A similar parameter can be introduced for the effective magnetic field $\mathbf{H}^\mathrm{IFE}$ so that mode volume can be calculated as $V_\mathrm{mode}=\int (\max|H^\mathrm{IFE}|)^{-1}|\mathbf{H}^\mathrm{IFE}(\mathbf{r})|\mathrm{d}\mathbf{r} $. Mode volumes can be introduced for each of the $\mathbf{H}^\mathrm{IFE}$ components in the same manner. Table~\ref{tab:Mode_volume} summarizes the calculated $V_\mathrm{mode}$ for each of the Mie mode under consideration. 

\begin{table*}[]
    \centering
    \begin{tabular}{c||c|c|c|c|c|c}
 \hline
 \(H^\mathrm{IFE}\) component & \(V_\mathrm{MD}/V_\mathrm{sphere}\) & \(V_\mathrm{ED}/V_\mathrm{sphere}\) & \(V_\mathrm{MQ}/V_\mathrm{sphere}\)& \(V_\mathrm{EQ}/V_\mathrm{sphere}\)& \(V_\mathrm{MO}/V_\mathrm{sphere}\) & \(V_\mathrm{EO}/V_\mathrm{sphere}\)\\
 \hline
 \(H^\mathrm{IFE}_x\) & 0.3552 & 0.2925 & 0.3075 & 0.1194 & 0.2358 & 0.0776\\
 \(H^\mathrm{IFE}_y\) & 0.3552 & 0.2925 & 0.3075 & 0.1194 & 0.2358 & 0.0776\\
 \(H^\mathrm{IFE}_z\) & 0.2806 & 0.0746 & 0.1254 & 0.1224 & 0.0746 & 0.0985\\
 \hline
 \(|\mathbf{H^\mathrm{IFE}}|\) & 0.4209 & 0.0925 & 0.197 & 0.1731 & 0.1075 & 0.1373\\

 \hline
    \end{tabular}
    \caption{Ratio of the mode and sphere volumes $V_\mathrm{mode}/V_\mathrm{sphere}$ for absolute value and components of $\mathbf{H}^{IFE}$ at MD, ED, MQ, EQ, MO, EO Mie resonances of the nanosphere.}
    \label{tab:Mode_volume}
\end{table*}

The lowest order Mie mode, namely MD, has the largest volume $V_\mathrm{mode}$. This mode occupies more than $40 \%$ of the nanosphere volume ($V_\mathrm{sphere} = 0.0335 \mu m^3$). With the increase of the magnetic mode order (MD, MQ, MO), the mode volume rapidly decreases. At the same time, the \(H^\mathrm{IFE}_{z}\) distributions of all of the considered magnetic modes are similar to each other (Fig.~\ref{fig: IFE_iso}). \(H^\mathrm{IFE}_{z}\) field is concentrated in the two hotspots located at the top and bottom parts of the nanosphere.

The electric family (ED, EQ, EO) of the Mie modes has nearly identical sum mode volumes. At the same time, one can observe how the spatial distribution changes and the number of \(H^\mathrm{IFE}_{z}\) hotspots increases (1 for ED, 2 for EQ, and 4 for EO) with the resonance order increase.  

Such a different behaviour of the mode volume and field distribution for the electric and magnetic resonance families make Mie-supporting nanoparticles interesting from the practical point of view. On the one hand, large mode volumes are important for efficient optomagnonic coupling between magnons and light within optical resonators~\cite{rameshti2022cavity}. From this perspective, the magnetic dipole mode appears to be very promising. On the other hand, small mode volumes and tunable localization of the IFE field hotspots are important for the excitation of the standing spin modes of different orders~\cite{ignatyeva2024optical}. Switching the excitation wavelength in a quite moderate range from 750~nm to 550~nm one could excite different electric resonances having drastically different number and location of the IFE hotspots. Thus, by tuning the laser pump to excite the electric resonances, one can control the number of the spin-wave sources within a nanoparticle.

\section{Sign-changing IFE inside a nanoparticle}
Figure~\ref{fig: IFE_iso} depicts that the IFE field is distributed inhomogeneously inside the nanoparticle. While $H^{IFE}_z$ is primarily localized at the poles or in the center of the sphere, $H^{IFE}_{x,y}$ is distributed throughout the entire volume. In contrast to $H^\mathrm{IFE}_{z}$, whose sign is determined by light helicity, the two other components ($H^\mathrm{IFE}_{x}$ and  $H^\mathrm{IFE}_{y}$) are odd with respect to the $z=0$ plane. Moreover, $H^\mathrm{IFE}_{x}$ and $H^\mathrm{IFE}_{y}$ are odd with respect to the $x=0$ and $y=0$ planes, respectively. Thus, $x$ and $y$ components of IFE possess sign-changing profiles with a zero average values. 

One might expect that such IFE magnetic field distributions could excite inhomogeneous standing spin modes of a nanosphere. While rigorous theoretical analysis of spin dynamics in such systems is quite complicated \cite{arias2005dipole} and beyond of the present research, let us concentrate on the differences in the inhomogeneous patterns of the optically-induced IFE magnetic fields provided by different orders of Mie resonances. The impact of the sign-changing IFE distributions on the excited spin dynamics is discussed qualitatively.

\begin{figure*}[htbp]
\centering
\includegraphics[width=1\textwidth]{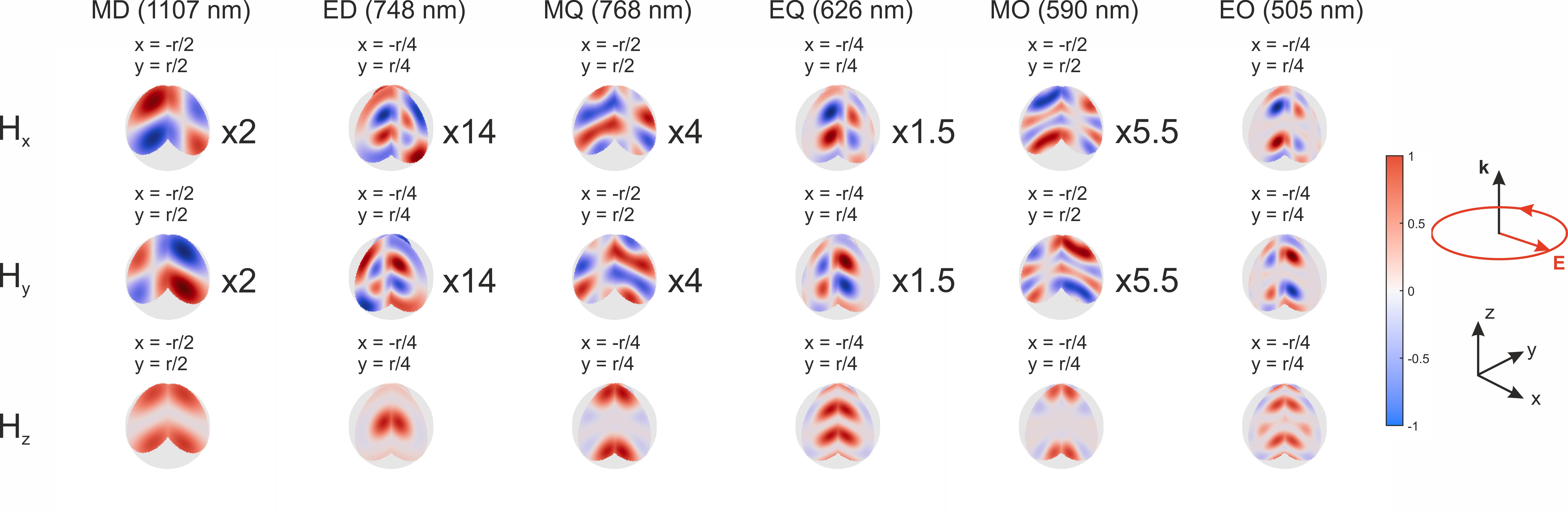} 
\caption{Crossections of $H^{IFE}$ fields at resonant wavelengths corresponding to the Mie modes excitation. The planes at which the cross-sections are taken (see the legends) are selected to show the areas with maximal magnitude of IFE. The numbers indicate the ratio between the $H^\mathrm{IFE}_{x}$, $H^\mathrm{IFE}_{y}$ and $H^\mathrm{IFE}_{z}$ components at the fixed wavelength.}
\label{fig: IFE_slices}
\end{figure*}

To analyze the \(H^\mathrm{IFE}_{x,y,z}\) patterns, it is more convenient to look closely at the slices of these distributions (see Fig.~\ref{fig: IFE_slices}). One can clearly see how the number of nodes in $\mathbf{H}^\mathrm{IFE}$ distribution varies for the Mie resonances of different orders. The distributions of \(H^\mathrm{IFE}_x\) and \(H^\mathrm{IFE}_y\) components are identical, with exception of a $\pi/2$-angle spatial rotation around $z$-axis. Such similarity of \(H^{IFE}_x\) and \(H^{IFE}_y\) functions is observed for all Mie resonances. This is due to the considered system spatial rotational symmetry being excited by a circularly polarized light, which can be represented as the sum of the two orthogonal linearly polarized components with a $\pi/2$ phase shift. Thus, only \(H^\mathrm{IFE}_x\) and \(H^\mathrm{IFE}_z\) will be discussed in detail.

Let us have a closer look at the magnetic family of the Mie resonances (see Fig.~\ref{fig: IFE_slices}). For MD resonance the period of \(H^\mathrm{IFE}_{x}\) is nearly equal to nanosphere diameter (400 nm) along z- and x-axis. One might expect that such a distribution could efficiently excite the standing spin mode with the same wavelength and $N_{x,z}=1$ one node inside a nanoparticle in $x$ and $z$ directions. For MQ resonance the IFE field forms a standing-wave-like distribution along $x$ and $z$ directions with periodicity close equal to the nanosphere radius (200 nm). MO mode has similar to MQ distribution, but it is slightly compressed from the poles. As a result, for MQ and MO modes one might expect excitation of the standing spin waves with $N_{x}=1$ and $N_{z}=3$ nodes inside a nanoparticle. For all magnetic type resonances \(H^\mathrm{IFE}_z\) is found to be concentrated at the poles of the nanosphere. 

For the electric family of modes (see Fig.~\ref{fig: IFE_slices}) the number of the nodes of \(H^\mathrm{IFE}_x\) function is nearly the same for ED, EQ and EO modes. Furthermore, there are three nodes of this function in $x$, $y$ and $z$ directions. One might expect that such modes would excite standing spin waves at least with $N_{x,y,z}=3$. At the same time, for EQ mode the IFE maxima are shifted to the poles so that the effective period of the sign-changing distribution is diminished to one third of a sphere diameter. Thus, even higher spin modes with $N_{x,y,z}=5$ nodes can be excited.

\section{Vortex IFE structure under Mie mode excitation}

\begin{figure*}[htbp]
\centering
\includegraphics[width=0.9\textwidth]{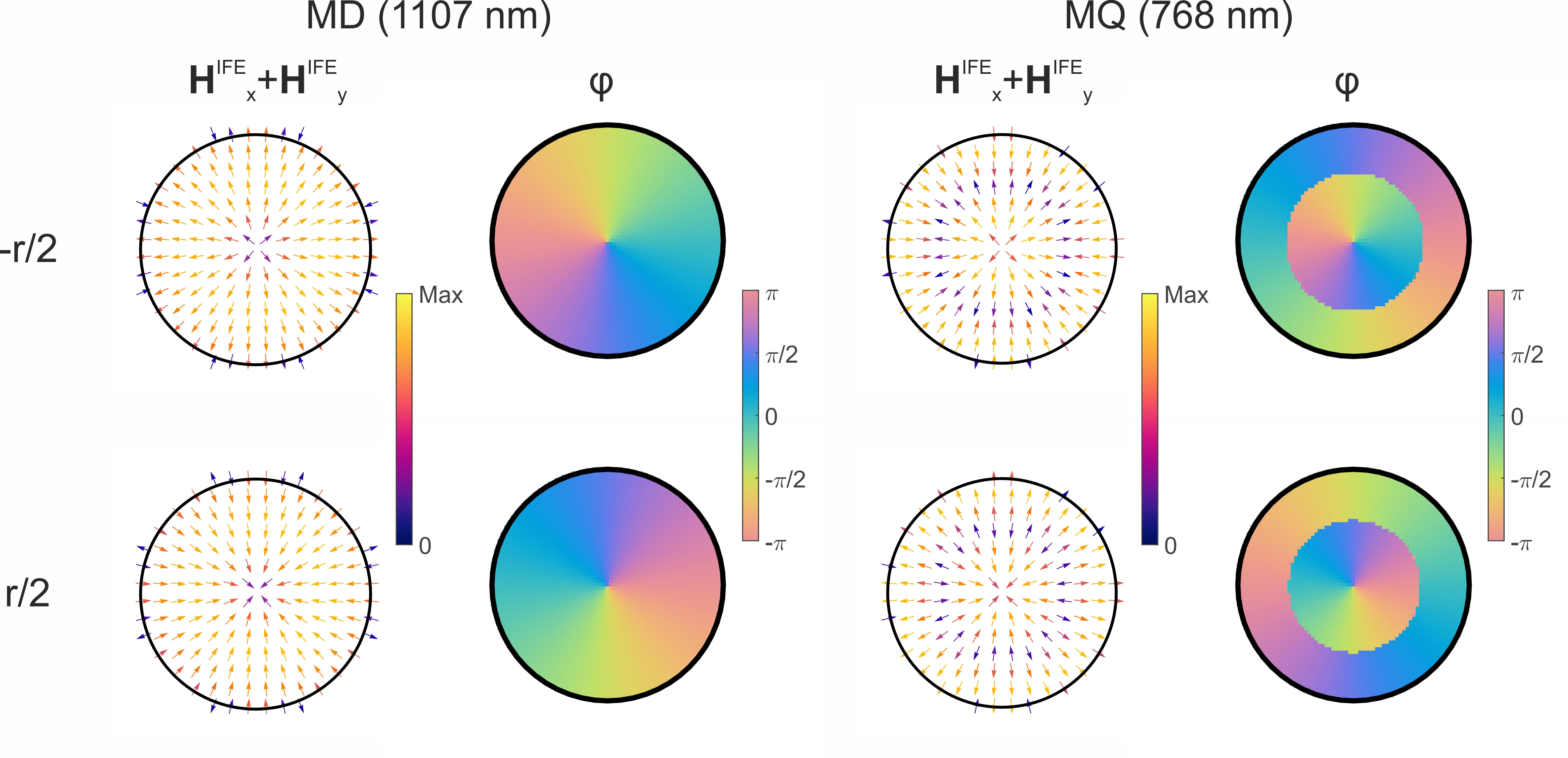} 
\caption{ $\mathbf{H}^{IFE}_x+\mathbf{H}^{IFE}_y$ and its spatial phase (azimuthal angle) $\varphi$ distribution at MD and MQ resonances for $z=\pm r/2$. The magnitude of the field is shown by color in vector plots.}
\label{fig: IFE_vectors}
\end{figure*}

For optomagnonics it is crucial to know direction of the effective field which launches spin-wave dynamics. Thus, it is important to analyze not only the components, but the entire vector field $\textbf{H}^\mathrm{IFE}=\textbf{H}^\mathrm{IFE}_x+\textbf{H}^\mathrm{IFE}_y+\textbf{H}^\mathrm{IFE}_z$ to determine the direction of the generated effective fields at each resonance. Although $\textbf{H}^\mathrm{IFE}_z$ dominates in magnitude for all of the optical resonances, some peculiar features may emerge due to the special distributions of $\mathbf{H}^\mathrm{IFE}_{x,y}$ fields. If the external magnetic field saturate the magnetic particle along $z$-axis, $\textbf{H}^\mathrm{IFE}_z$ component of the induced magnetic field does not produce any torque to the spin system and the dynamics is solely driven by $x-y$ plane components only.  Figure~\ref{fig: IFE_vectors} shows the $\textbf{H}^\mathrm{IFE}$ field $x-y$ plane components at the several cross-section of the nanosphere. 

An interesting feature of Mie resonance-induced magnetic fields is that $\textbf{H}^\mathrm{IFE}_{xy}=\textbf{H}^\mathrm{IFE}_{x}+\textbf{H}^\mathrm{IFE}_{y}$ possesses vortex behaviour at each resonances. This is confirmed by $2\pi$ twist of the magnetic field direction $\varphi = \tan^{-1} ({H}^\mathrm{IFE}_y/{H}^\mathrm{IFE}_x)$ (see Fig.~\ref{fig: IFE_vectors} for MD and MQ modes and Supplemental Material for the other resonances).  All of these vortices resemble so-called "spin skyrmions" which are static optical skyrmions \cite{shen2023optical}. Mie-mode induced  $\textbf{H}^\mathrm{IFE}_{xy}$ fields belong to the Neel type of optical skyrmions, with radial components are directed outwards or inwards from the circle center (see Figure \ref{fig: IFE_vectors}). 

Skyrmions are characterized by 3 main topological numbers: polarity \textit{p}, vorticity \textit{m}, and the helicity \textit{$\gamma$}. The polarity of the vortex \textit{p} defines if $\mathbf{H}_z^{IFE}$ component is reversed at the boundary of nanosphere comparing its center. For the external magnetic field applied along $z$-axis the induced $\mathbf{H}_z^{IFE}$ does not influence spin dynamics, so the polarity of the IFE-field induced magnetic structure is zero, $p=0$. 

Vorticity \textit{m} is defined by distribution of the in-plane components of the field and indicate the number of $2\pi$ rotations of $\textbf{H}^\mathrm{IFE}_{xy}$ around the center of the sphere. The vorticity $m=1$ for all of the Mie-mode induced IFE fields (Fig.~\ref{fig: IFE_vectors}). This feature is also due to the rotational symmetry of the given configuration.

The helicity of the vortex is determined by an initial phase \textit{$\gamma$}, which contributes to total phase defined by \textit{m}~\cite{shen2023optical}. This means that the vector fields should point outwards for $\gamma=0$ and inwards for $\gamma=\pi$. It is interesting that the top and bottom parts of the nanosphere nest the optically-induced $\textbf{H}^{IFE}_{xy}$ skyrmions with the opposite $\gamma=0$ and $\gamma=\pi$, respectively (see the cross-sections for $z=r/2$ and $z=-r/2$ in Fig.~\ref{fig: IFE_vectors}). Inward or outward direction of $\textbf{H}^{IFE}_{xy}$ is determined by the relative phase of $E_{xy}$ and $E_z$ light electric field components, which, in their turn are controlled by the helicity (right or left circular polarization) of the incident light. As a result, changing the light helicity causes the IFE-induced skyrmion's helicity to switch. At MD resonance  $\textbf{H}^{IFE}_{xy}$ skyrmion the topological numbers are $\textit{p}=0$, $\textit{m}=1$ and \textit{$\gamma$}=0 or $\pi$ (Fig.~\ref{fig: IFE_vectors}).

At the same time, one might see that several IFE distributions in Fig.~\ref{fig: IFE_slices}, especially the MQ and EQ modes, are more complex and possess sign-changing in radial directions $\textbf{H}^{IFE}_{xy}$ components. This results in a very peculiar double-ring pattern (Fig.~\ref{fig: IFE_vectors}b). The inner and outer parts of the skyrmion are characterized by the opposite spatial phase and different helicities, namely \textit{$\gamma$}=0 or $\pi$. The situation is the turns upside down for the top and the bottom parts of the nanosphere, and, again, can be switched to the opposite one by the light helicity change.

Thus, optically-induced fields possess a complex vortex structure inside the nanosphere due to the excitation of the Mie rosonances. The topology of the IFE field skyrmion structures can be controlled by tuning the laser pump helicity and wavelength.

\begin{figure*}[htbp]
\centering
\includegraphics[width=0.9\textwidth]{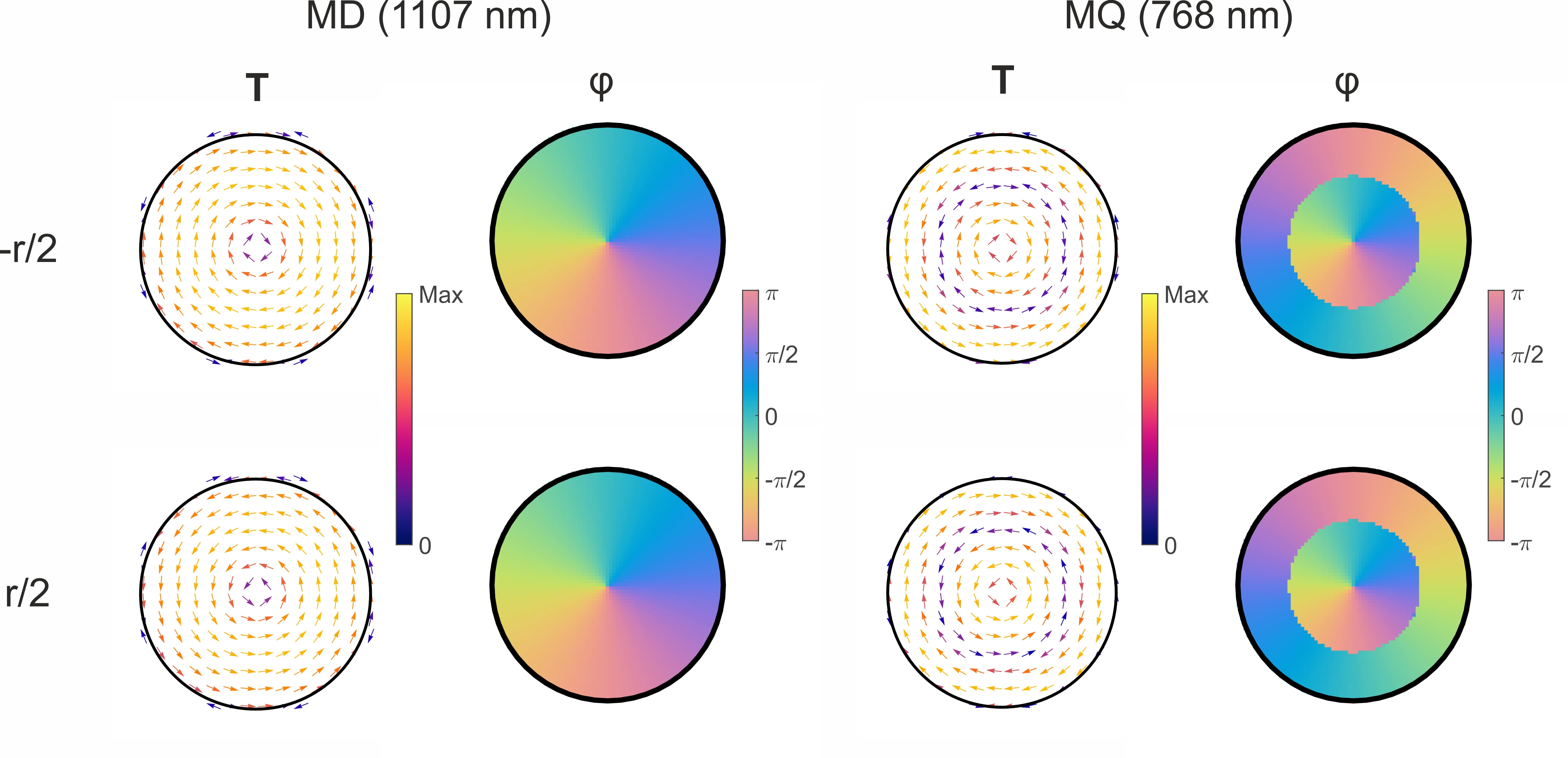} 
\caption{ IFE-induced torque $\mathbf{T}$ and its spatial phase (azimuthal angle) $\varphi$ distribution at MD and MQ resonances for $z=\pm r/2$. The magnitude of the field is shown by color in vector plots.}
\label{fig: torque_vectors}
\end{figure*}

IFE is a driving force that launches spin dynamics via a toque defined by cross product $\mathbf{T}\propto\mathbf{M}\times\mathbf{H}^\mathrm{IFE}$ ~\cite{landau1992theory}. We calculated normalized value of $\mathbf{T}=\mathbf{M}\times \mathbf{H}^\mathrm{IFE}/(|\mathbf{M}|\max(|\mathbf{H}^\mathrm{IFE}|))$ to figure out its direction. As the nanosphere uniformly magnetized along OZ direction is considered, only $\mathbf{H}^\mathrm{IFE}_{xy}$ components induce dynamics. The calculated $\mathbf{T}$ distribution for MD and MQ resonances is presented in Fig. ~\ref{fig: torque_vectors} . In contrast to vortices observed for $\mathbf{H}^\mathrm{IFE}_{xy}$, $\mathbf{T}$ has different topological number $\gamma=\pm \pi/2$ which result in formation of Bloch type vortex ~\cite{shen2023optical}. This feature is also clearly seen on the phase distribution (Fig.~\ref{fig: torque_vectors}) which acquire additive $-\pi/2$. Notice that in the case of the IFE-excited precession the helicity of the vortex will change with a precession frequency.

\section{Outlook}

The observed features of the Mie-mode induced IFE fields are particularly intriguing for advanced optomagnonic devices. In optomagnonics short laser pulse with subpicosecond duration ($\sim$5-500 fs depending on the laser system) is utilized to excite magnetization dynamics via generated effective fields at the same timescales. In bulk materials the generated $\textbf{H}^\mathrm{IFE}$ is uniformly distributed on a micrometer spatial scales. By nanostructuring the material and pumping at the Mie resonance wavelength, it is possible to create a nanometer-scaled inhomogeneous spin dynamics source with variable characteristics depending on the resonance type. An important feature of this system is that Mie resonances are spectrally detuned and can usually be excited separately by proper selection of pump wavelength, which gives spectral tunability to the system.  
\par Excitation of electric family of the Mie resonances enables one to excite spin dynamics by one, two, or four point-like sources with a volume less than 1/10 of the nanosphere volume each. It was recently demonstrated that a small source of the spin dynamics inside a nanoparticle launches several standing spin waves whose profiles are determined by the excitation field size~\cite{ignatyeva2024optical}. An interesting possibility is that one may switch between different excitation regimes in Mie-resonant particles by changing the pump wavelength from $\sim750$~nm (ED resonance, one spot) to $\sim505$~nm (EO resonance, four hotspots). 
\par Magnetic family of the Mie resonances is characterized by much larger volumes of $\textbf{H}^\mathrm{IFE}$ field distributed in a similar way for all of the MD, MQ, and MO modes. This improves the interaction efficiency of optomagnonic coupling between the resonant optical photons and the magnons, for example, of a Kittel mode. This may significantly miniaturize down to a nanometer scale the light-magnon coupling devices, which are typically represented by optical microresonators~\cite{bittencourt2019magnon,graf2018cavity,parvini2020antiferromagnetic, almpanis2020spherical,almpanis2018dielectric,almpanis2021nonspherical} with millimeter characteristic dimensions.
\par While the light wavevector determines the direction of the IFE field in a bulk material, Mie-supporting nanoparticles enables optically induced inhomogeneous and periodically sign-changing $\textbf{H}^\mathrm{IFE}_{x,y}$ fields. Thus, as the standing spin excitation efficiency depends on whether the distribution of the excitation magnetic filed coincides the spin eigen mode spatial distribution, Mie-suporting nanoparticles provide the possibility to launch spin waves modes with higher selectivity. One might expect that the typical wavelengths of the excited standing spin waves should be around 400~nm for MD mode, 200~nm for ED, MQ, EQ and EO resonances and nearly 75~nm for MO. Similar standing-wave like behaviour of the IFE field can launch standing spin waves along all of the three spatial dimensions.

Magnetic skyrmions seem to be a promising candidate for the magnetic information writing and processing technologies. It is an important feature of the Mie-resonance-induced effective magnetic fields that $\textbf{H}^{IFE}_{xy}$ components possess the vortex structure. Topology of the IFE-generated fields changes with variation of the optical pump wavelength due to the change of the resonance type. Moreover, the excitation laser pulse's helicity can be used to control the helicity of the skyrmion pattern. Spectral position of Mie resonances can be controlled by different shapes of the nanoelement \cite{verre2019transition,wu2014spectrally, li2016all} which can be produced by electron or ion-beam lithography \cite{baranov2017all}. This makes Mie-resonance nanoparticles quite promising for magnetic skyrmions generation which are predicted to exist in a submicron magnetic particles \cite{pathak2021three,baral2022tuning,stavrou2021magnetic,li2020stabilization} of a different shapes and generated by optical vortices in smooth films \cite{fujita2017ultrafast}.\\ 

\section{Conclusion}
In this study we demostrated theoretically and by numerical simulation that optical Mie resonances in magnetic BIG nanoshpere can cause complex behaviour of the IFE effective fields. The IFE nonuniform alternating-sign distribution at Mie resonances can be used for standing spin waves generation inside nanopartices of a different forms and shapes. The order of standing spin modes can be adjusted by sweeping laser wavelength. Modes with large effective magnetic field volumes may be potentially interesting for optomagnonic coupling at the nanoscale. The observed inhomogeneities give rise to skyrmion patterns of $\textbf{H}^{IFE}_{x,y}$ fields whose topological numbers can also be controlled by the laser pump wavelength and helicity. Our approach paves the way for advanced light-magnon interaction in iron garnet nanoparticles and advances the development of controllable all-optical optomagnonic devices.

\section*{acknowledgement}

This work was financially supported by the Russian Science Foundation, project No. 21-72-10020.



\bibliography{bibliography}

\end{document}


\title{Supplemental Material for "The inverse Faraday effect at Mie resonances"}
\preprint{APS/123-QED}

\author{Denis M. Krichevsky}
\email{krichevskii.dm@phystech.edu}
\affiliation{Russian Quantum Center, 121205 Moscow, Russia}
\affiliation{Moscow Institute of Physics and Technology (National Research University), Dolgoprudny 141700, Russia}

\author{Daria O. Ignatyeva}
\email{daria.ignatyeva@gmail.com}
\affiliation{Russian Quantum Center, 121205 Moscow, Russia}
\affiliation{Faculty of Physics, M.V. Lomonosov Moscow State University, 119991 Moscow, Russia}

\author{Vladimir I. Belotelov}
\affiliation{Faculty of Physics, M.V. Lomonosov Moscow State University, 119991 Moscow, Russia}
\affiliation{Russian Quantum Center, 121205 Moscow, Russia}


\date{\today}

\maketitle

\section{Section S1. Vortex IFE structure of ED, EQ, MO, EO Mie resonances.}

\begin{figure*}[htbp]
\centering
\includegraphics[width=1\textwidth]{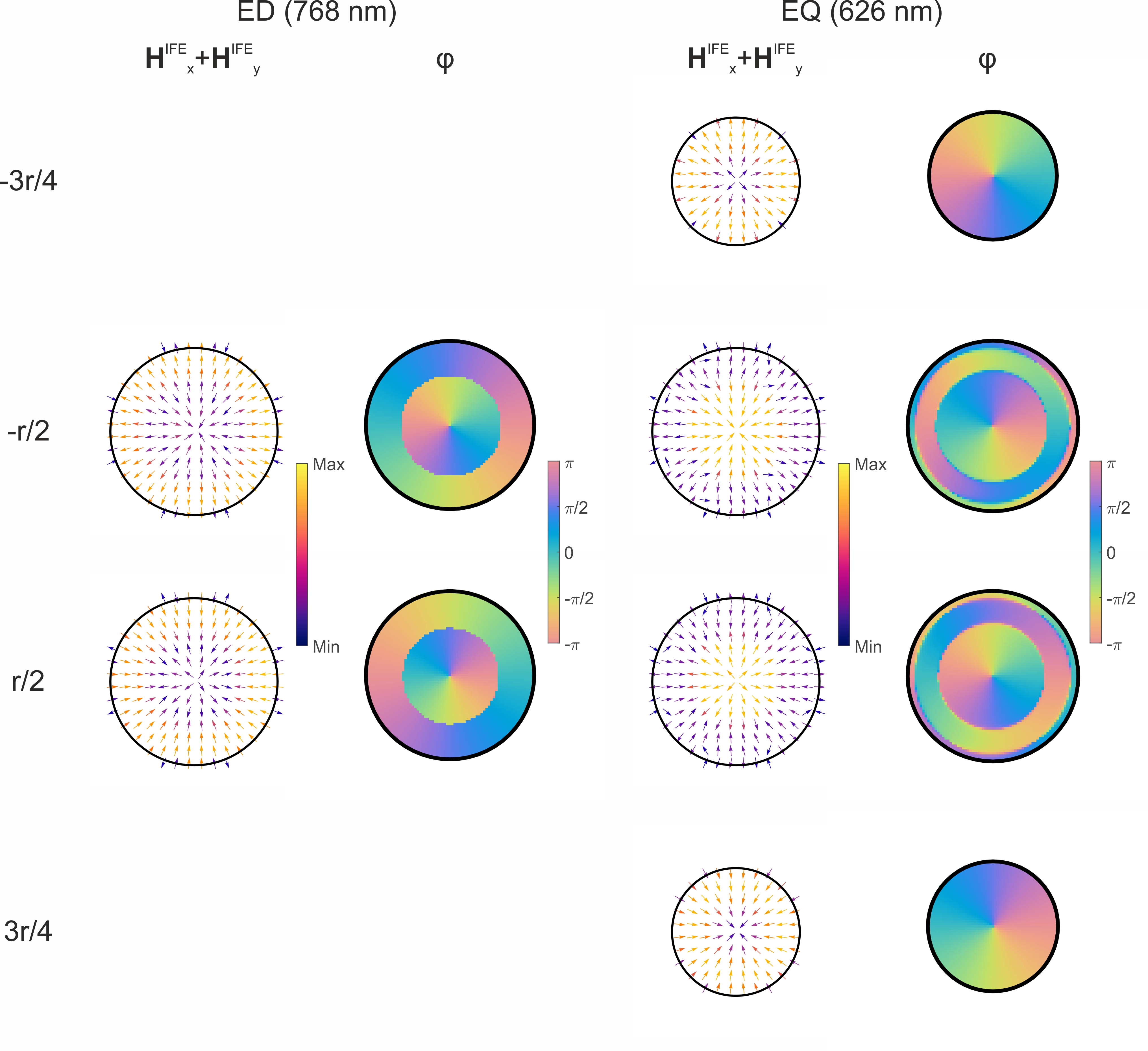} 
\caption{$\mathbf{H}^{IFE}_x+\mathbf{H}^{IFE}_y$ and spatial phase $\varphi$ distribution at ED and EQ resonances}
\label{fig: IFE_vectors}
\end{figure*}

\begin{figure*}[htbp]
\centering
\includegraphics[width=1\textwidth]{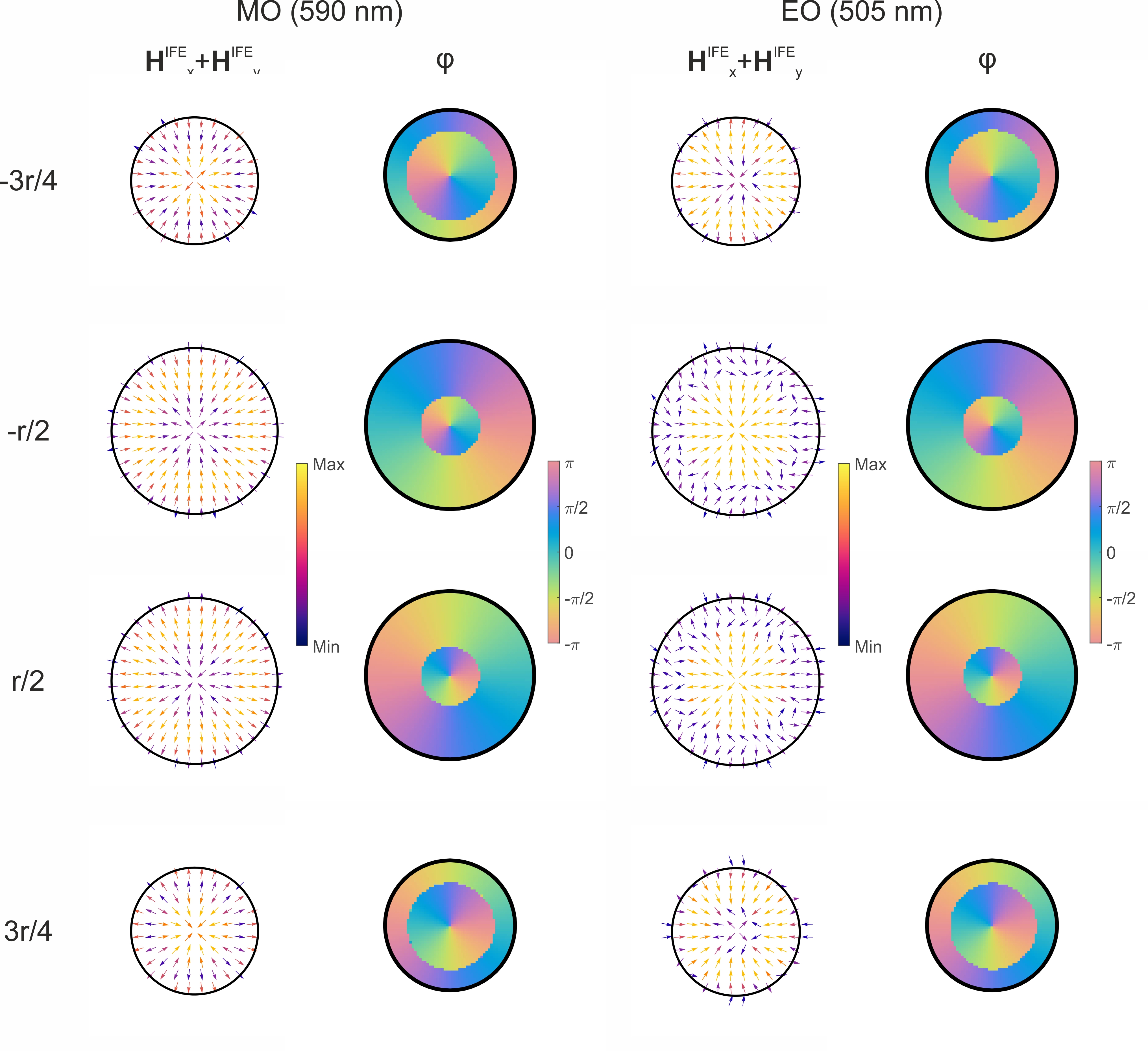} 
\caption{$\mathbf{H}^{IFE}_x+\mathbf{H}^{IFE}_y$ and spatial phase $\varphi$ distribution at MO and EO resonances}
\label{fig: IFE_vectors_1}
\end{figure*}

\bibliography{bibliography}